\documentclass[aps,prl,twocolumn,showpacs,a4paper,superscriptaddress]{revtex4-1}
\usepackage{graphicx,bm,amsmath,dcolumn,amssymb}
\usepackage{longtable,mathtools}
\usepackage{newtxtext,newtxmath}
\graphicspath{{figures}}
\usepackage{geometry,hyperref,color}
\begin{document}
\title{Universal Quench Dynamics of an Open Quantum System}
   \author{Chengxiang Ding}
  \email{dingcx@ahut.edu.cn}
   \affiliation{School of Microelectronics $\&$ Data Science, Anhui University of Technology, Maanshan 243002, China }
   \author{Long Zhang}
   \affiliation{Kavli Institute for Theoretical Sciences and CAS Center for Excellence in Topological Quantum Computation, University of Chinese Academy of Sciences, Beijing 100190, China}
   	\date{\today}
\begin{abstract} 
Taking the quantum Kitaev chain as an example, we have studied the universal dynamical behaviors resulting from quantum criticality under the condition of environmental temperature quench. Our findings reveal that when the quantum parameter is at its critical value, both the excess excitation density at the end of linear quench and the subsequent free relaxation behavior exhibit universal scaling behaviors. The scaling laws observed upon quenching to the zero-temperature quantum critical point and non-zero temperature points exhibit distinct scaling exponents, which are all intimately related to the dynamical critical exponents of the quantum phase transition.
Additionally, for the case of linear quench to finite temperatures, we have also discovered an intrinsic universal dynamical behavior that is independent of quantum criticality.
Our research offers profound insights into the relationship between quantum criticality and  nonequilibrium dynamics from two perspectives: Kibble-Zurek-like scaling behavior and free relaxation dynamics. 
 Notably, the Kibble-Zurek-like scaling behavior in this context differs from the standard Kibble-Zurek mechanism. These two aspects jointly open up a new avenue for us to understand quantum criticality through real-time dynamical behavior, even at finite temperatures.
\end{abstract}
\maketitle 

{\it Introduction.}---The concept of universality, developed from the study of equilibrium phase transition and critical phenomena, is not only applied to equilibrium physics, but also widely used in nonequilibrium physics.
Many dynamical processes also exhibit universal scaling laws, which can be referred to as ``dynamic universality." 
One typical example is the Kibble-Zurek mechanism\cite{KZ1,KZ2,KZ3,KZ4}, which focuses on the universal  scaling law between the topological excitation number of the system and the driving rate after the system undergoes a linear or nonlinear driving through a phase transition point. 
Related conclusions are universally applicable to both classical and quantum phase transitions\cite{KZ3,KZ4}. So far, the Kibble-Zurek mechanism has been extensively studied, verified and expanded in various systems\cite{kze1,kze2,kze3,kze4,kze5,kze6}, and verified by experiments\cite{KZexp}.
Currently, new phenomena and theories related to this question are still emerging, such as the Kibble-Zurek mechanism across nonequilibrium phase transitions\cite{neqKZ}, the explanation of topological defect distributions by large deviation theory\cite{bigDev}, the ``anti-Kibble-Zurek mechanism"   in noisy systems\cite{noisyKZ}, 
the universal breakdown of Kibble-Zurek scaling in fast quenches\cite{fastQC}, the Kibble-Zurek scaling in the Yang-Lee edge singularity\cite{YLKZ}, and so forth.

Unlike the Kibble-Zurek mechanism, another focus of research on dynamical universality is the real-time dynamic behavior of certain physical quantities during nonequilibrium processes. These processes can be either a driving process or a free relaxation process that follows a quench. Typical examples of relaxation process research include phase ordering and aging phenomena. Phase ordering focuses on the universal laws governing the growth of the characteristic length of the system over time after quenching to an ordered phase\cite{pod,podlong,podIsing,podIisng2}, while aging focuses on the autocorrelation characteristics of the order parameter before and after the quench\cite{ag1,ag2}. Research on these two types of questions has been extended to random field models\cite{rand-ag}, nonequilibrium lattice gas models\cite{lgas-ag}, polymer folding problems\cite{pol-ag},  aging problems starting from critical points\cite{cri-ag}, and so forth. Currently, the two questions remain hot topics in nonequilibrium statistical physics. Relaxation process research can also be extended to isolated quantum systems, although the stable state ultimately reached by the system after the quench is not necessarily a thermalized state that follows the Gibbs distribution. Recent research has shown that in the relaxation process after a sudden quench, the transient value of the system's short-range correlation function approaching the stable value is also a decaying process with universal power law, and the decaying exponent depends only on whether the system is quenched to a commensurate or incommensurate phase, regardless of the initial state of the quench\cite{iso1,iso2}, unless the initial state is a critical point of equilibrium phase transition\cite{I-2023,I-2024,C-2024}. 
Recently, the relaxation process of isolated quantum  system after linear quench has also garnered attention,  it is shown that
 the post-quench state is a superposition of distinct, broken-symmetry vacua with different numbers and locations of defects,
 and this non-classical characteristic may lead to coherent quantum oscillations exhibiting universal laws\cite{osci,osci2}.

Extending the concept of dynamical universality, particularly the Kibble-Zurek mechanism, to open quantum systems has emerged as a hot topic in recent years, prompting significant and groundbreaking research efforts\cite{SD,noiseD,open1,open2,open3,open4,open5,bath1,bath2,bath3,bath4,bath5}. One intriguing aspect of this endeavor involves placing the system within a Markovian thermal bath and examining the Kibble-Zurek scaling behavior induced by changes in the bath or system parameters\cite{bath1, bath2, bath3,bath4,bath5}. For instance, recent studies have uncovered that when the environmental temperature is linearly cooled towards a quantum critical point, the excitation density of the system adheres to the predictions of the Kibble-Zurek mechanism\cite{bath4,bath5}. However, there has been limited exploration, to the best of our knowledge, on the free relaxation behavior subsequent to such cooling and the dynamical behaviors upon quenching to finite temperatures, including the scaling behaviors akin to the Kibble-Zurek scaling and the post-quench free relaxation dynamics. In this paper, we delve into this underexplored territory.

\begin{figure}[thpb]
	\centering
	\includegraphics[width=0.75\columnwidth]{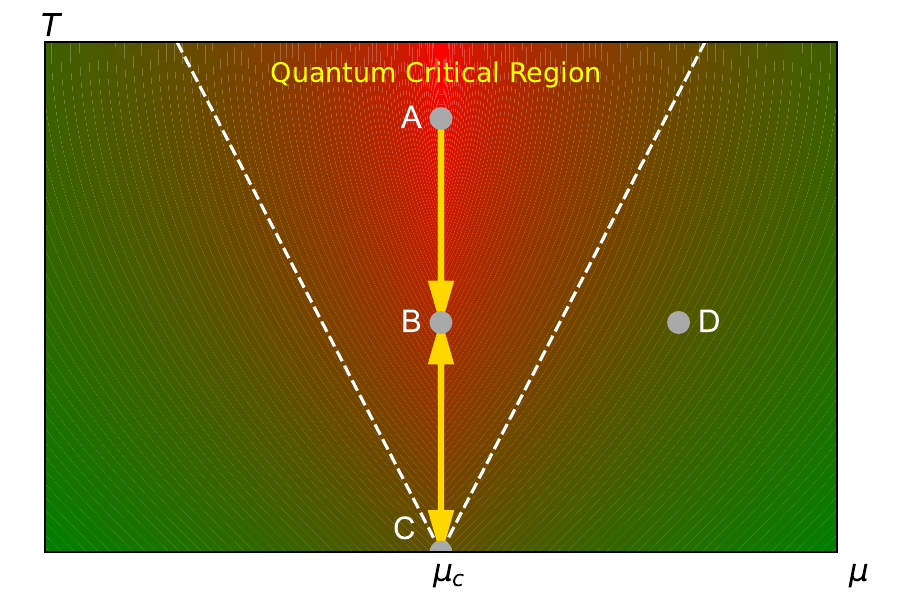}
	\caption{(Color online) Quenching paths of model (\ref{shortfermion}): For linear quenches, we only consider the paths along AC, where $\mu=\mu_c=1$; for sudden quenches, we also consider the case where $\mu\ne \mu_c$, such as a sudden quench from D to B or C. 
    In this paper, if not explicitly stated, the value of $\chi$ is taken as 1.  }
	\label{pd}
\end{figure}

For clarity, Fig. \ref{pd} outlines the quantum Kitaev chain model and various quench paths that we will examine. 
We found that if the environmental temperature is linearly quenched along the critical line AC, the excess excitation density at the end of the quench will exhibit Kibble-Zurek-like scaling behavior; as long as the endpoint of the quench lies on the AC line, the relaxation behavior after the quench also displays universal scaling behavior. For these two issues, the scaling laws exhibited by quenching to the quantum critical point and other nonzero-temperature points on the critical line AC possess different power exponents, which are closely related to the dynamical critical exponents of quantum phase transitions.
Additionally, in the case of linear quench to finite temperatures, we have also discovered an intrinsic universal dynamical behavior that is independent of quantum criticality, namely, scaling laws that exist even when $\mu$ is not equal to $\mu_c$.

{\it Model and Method.}---We study a quantum Kitaev chain 
\begin{eqnarray}
	H=\sum\limits_{j=1}^L(c_j^\dagger c_{j+1} +\chi c_j^\dagger c_{j+1}^\dagger+{\rm H.c.})-2\mu\sum\limits_{j=1}^L c_j^\dagger c_j, \label{shortfermion}
\end{eqnarray}
The model is in an ordered phase when $|\mu|<1$  and in a disordered phase when $|\mu|>1$; $(\mu=1, \chi\ne1)$ represents the critical line with a dynamical critical exponent of $z=1$, while $(\mu=1, \chi=1)$ is a multicritical point with a dynamical critical exponent of $z=2$.  In this paper, if not explicitly stated, we set $\chi=1$. 

The model is coupled  to a thermal bath with the Lindblad form\cite{lindblad}
\begin{eqnarray}
\dot{\rho}=- i[H,\rho]+\sum\limits_k\sum\limits_{\sigma=\pm}\gamma_{k\sigma}\Big(2L_{k\sigma}\rho L_{k\sigma}^\dagger-\{L_{k\sigma}^\dagger L_{k\sigma},\rho\}\Big),
\end{eqnarray}
where $\rho$ is the density matrix operator of the Kitaev chain,  the jump operators $L_{k+}=\eta_k^\dagger$ and  $L_{k-}=\eta_k$, 
with $\eta_k^\dagger$ and $\eta_k$ the fermionic creation and annihilation operators of the quasiparticle of the Kitaev chain, 
i.e., $H=\sum_k\varepsilon_k\eta_k^\dagger\eta_k$, where $\varepsilon_k=2[(\mu-\cos k)^2+\chi^2\sin^2k]^{1/2}$ is the energy spectrum.
  The information of the thermal bath are contained in  $\gamma_{k\sigma}$ , where 
\begin{eqnarray}
\gamma_{k+}= \mathcal{S}(\varepsilon_k) f(\varepsilon_k), \quad \gamma_{k-}= \mathcal{S}(\varepsilon_k) [1-\zeta f(\varepsilon_k)].
\end{eqnarray}
Here  $f(\varepsilon_k)=1/(e^{\varepsilon_k/T}+\zeta)$ is the Fermi-Dirac distribution (if $\zeta$= +1) or the Bose-Einstein distribution (if $\zeta$= -1) of the bath;  $\mathcal{S}(\varepsilon_k)=\gamma_0\varepsilon_k^s$ contains the information of the  spectral density of the bath and control the coupling strength between the bath and the Kitaev chain. 

Our purpose is to study the quench dynamics of the model,  we pay special attention to the real-time dynamics of the excitation density  $\mathcal{D}(t)=\frac{1}{L}\sum_k\mathcal{P}_k(t)=\frac{1}{L}\sum_k\eta_k^\dagger\eta_k(t)$.  A general solution of this question and also the real-time dynamics of the other variables can be obtained by the method of Third Quantization\cite{ThirdQ1,ThirdQ2}.
Here we mainly study the case where quantum parameters remain unchanged, in this case $\mathcal{P}_k(t)=\eta_k^\dagger\eta_k(t)$ follows a very simple rate equation
\begin{eqnarray}
\frac{d\mathcal{P}_k(t)}{dt}=- 2(\gamma_{k+}+\gamma_{k-}) \big[\mathcal{P}_k(t)-\mathcal{P}_k^{\rm th}(\varepsilon_k/T)\big],\label{dynEQ}
\end{eqnarray}
where $\mathcal{P}_k^{\rm th}(\varepsilon_k/T)$ is the Fermion-Dirac distribution at temperature $T$.

\begin{figure}[thpb]
	\centering
	\includegraphics[width=0.75\columnwidth]{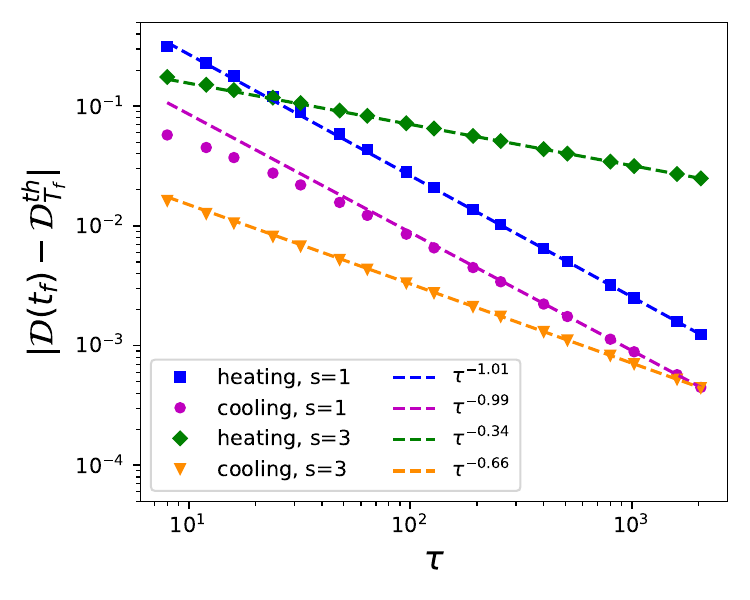}
	\caption{(Color online)  Kibble-Zurek-like scaling  behaviors of the Kitaev chain (\ref{shortfermion}) under linear 
		quench to a finite temperature with $(T_f,\mu)=(5,1)$; the starting point of heating is from the quantum critical point $(T_i,\mu)=(0,1)$, 
		and the starting point of cooling is from the point $(T_i,\mu)=(10,1)$. 
		The ramping protocol is $T(t)=T_i+t/\tau$ for the heating process and $T(t)=T_i-t/\tau$  for the cooling process, 
		$t_f=|T_i-T_f|\tau$ is the ending time of the quench.
		The other parameters are set as $\chi=1$, $\gamma_0=0.01$, and $L=10^4$.}
	\label{kzmfermi}
\end{figure}

{\it Kibble-Zurek-like scaling of linear quench}---
For the case of a linear cooling along the AC line depicted in Fig. \ref{pd} to the quantum critical point $(T,\mu)$=(0,1), the excitation density conforms to a scaling given by the standard Kibble-Zurek mechanism\cite{bath4}, which is written as
\begin{eqnarray}
	\mathcal{D}(t_f) \sim \tau^{-\frac{d}{z(s+1)}},\label{cooluc}
\end{eqnarray}
where $t_f$ is the ending time of the quench and $d=1$ is the dimension of the system. 
 This result holds true for quenches in both bosonic bath\cite{bath4} and fermionic bath\cite{bath5}. 
Regarding the relaxation process after such a quench, we will learn about it later. Here, we first focus on the Kibble-Zurek-like
scaling  for linear quenches to a finite temperature. 
For example, we consider a heating  from the quantum critical point $(T_i,\mu_i)$=(0,1) to a finite-temperature point $(T_f,\mu_f)$=(5,1);
 in a fermionic bath with Ohmic spectral density $\mathcal{S}(\varepsilon_k)=\gamma_0\varepsilon_k$,  
 we find that the excess excitation density  satisfy the scaling formula
$\mathcal{D}(T_f)-\mathcal{D}^{\rm th}_{T_f}\sim \tau^{-1}$, 
where $\mathcal{D}^{\rm th}_{T_f}$ is the excitation density of the thermal state at the final temperature $T_f$.
This is obviously different from the results of a cooling to the quantum critical point, where the scaling is $\tau^{-1/2}$,  given by Eq. (\ref{cooluc}).
Furthermore,  we find that for a quench from a high-temperature thermal state, we also get the  same scaling of $\tau^{-1}$;  these results are shown in Fig. \ref{kzmfermi}.
However,  the heating from the quantum critical point and the cooling from a high temperature do not always give a same scaling, it depends on the spectral density of the bath.
For example, for a quench in the fermionic bath with spectral density $\mathcal{S}(\varepsilon_k)=\gamma_0\varepsilon_k^3$,  
the scaling form is $\tau^{-1/3}$ for the heating process but $\tau^{-2/3}$ for the  cooling, this is also demonstrated in Fig. \ref{kzmfermi}.

These results can be understood by the solution of the rate equation (\ref{dynEQ}), for the case of fermion bath, it is written as 
\begin{eqnarray}
	\mathcal{P}_k(t)=\frac{e^{-2\gamma_kt_f}}{1+e^{\frac{\varepsilon_k}{T_i}}}+2\gamma_k\int_{0}^{t_f}\frac{e^{-2\gamma_k(t_f-t)}}{1+e^{\frac{\varepsilon_k}{T(t)}}}dt, \label{mainINT}
\end{eqnarray}
where $\gamma_k=\gamma_{k+}+\gamma_{k-}$.
For the case of a heating process from the quantum critical point with $(T_i,\mu)=(0,1)$,  the first term of the right hand of the equation is zero,  and the ramping protocol is $T(t)=T_i+t/\tau=t/\tau$,
thus $t_f=\tau T_f$, with $T_f$ the temperature of  the ending point of the quench. The spectral density of the fermionic bath is $\mathcal{S}(\varepsilon_k)=\gamma_0\varepsilon_k^s$; because it is at the quantum critical point, therefore $\varepsilon_k\sim k^z$, i.e., $\gamma_k\sim\gamma_0k^{zs}$, 
then the integration 
\begin{eqnarray}
	\mathcal{P}_k(t)\sim 2\gamma_0k^{zs}\int_{0}^{t_f}\frac{e^{-2\gamma_0k^{zs}(t_f-t)}}{1+e^{\frac{k^z}{t/\tau}}}dt.
\end{eqnarray}
Substituting  $k^{zs}(t_f-t)=t^\prime$ and taking the approximation $T_f-t^\prime/(\tau k^{zs})\approx T_f$, we get 
\begin{eqnarray}
	\mathcal{P}_k(t)&\sim &2\gamma_0\int_{0}^{t_fk^{zs}}\frac{e^{-2\gamma_0t^\prime}}{1+e^{k^z/T_f}}dt^\prime\\
	&=&\frac{1}{1+e^{k^z/T_f}}-\frac{e^{-2\gamma_0\tau T_fk^{zs}}}{1+e^{k^z/T_f}},
\end{eqnarray}
then integrating over $k$, we get
\begin{eqnarray}
	\mathcal{D}(t_f)-\mathcal{D}^{\rm th}(T_f) &&\sim - \frac{1}{\pi}\int \frac{e^{-2\gamma_0\tau T_fk^{zs}}}{1+e^{k^z/T_f}} dk\\
	&&=-\tau^{-\frac{1}{zs}}\cdot\frac{1}{\pi} \int \frac{e^{-2\gamma_0T_f k^{\prime zs}}}{1+e^{\frac{k^{\prime z}}{T_f\tau^{1/s}}}}dk^\prime \label{x1}\\
	&&\approx A(\tau)B\cdot\tau^{-\frac{1}{zs}} \sim \tau^{-\frac{1}{zs}}\label{x2}
\end{eqnarray}
where $\mathcal{D}^{\rm th}(T_f)=\frac{1}{\pi}\int dk/(1+e^{k^z/T_f}) $  is the excitation density of the thermal state at temperature $T_f$,  $k^\prime=k\tau^{1/(zs)}$, and  $A(\tau)=1/[ {1+\exp{(\frac{k^{\prime z}}{T_f\tau^{1/s}} )} }  ]$ is a function of $\tau$ but only weakly dependent on $\tau$,
because the contribution of the integration mainly comes from the vicinity of $k\rightarrow 0$. $B$ is a constant that does not depend on $\tau$. 
For the case $s=0$, the approximation $T_f-t^\prime/(\tau k^{zs})\approx T_f$ is not good enough, we need to make high order approximation, the specific processing steps  is very similar to the case of a cooling from a high temperature to a lower finite temperature\cite{SM}, the final result gives a scaling of  $\tau^{-1}$.

In summary, we conclude that for a heating from the quantum critical point to a finite temperature in a fermionic bath with spectral density $\mathcal{S}(\varepsilon_k) = \gamma_0\varepsilon_k^s$, the asymptotic behavior of the excess excitation density $\mathcal{D}(t_f) - \mathcal{D}^{\rm th}(T_f)$ at the end of a linear quench, when $\tau$ is sufficiently large, conforms to the following scaling
\begin{eqnarray}
	\mathcal{D}(t_f)-\mathcal{D}^{\rm th}(T_f) \sim 
		\begin{cases}  
		\tau^{-\frac{1}{zs}}, & zs\ge 1,\\  
		\tau^{-1}, &  zs<1. 
	\end{cases}  \label{fermikzm1}
\end{eqnarray}
Specially, we get a scaling of $\tau^{-1}$ for $s=1$ and $\tau^{-1/3}$ for $s=3$, as shown in Fig. \ref{kzmfermi}. More numerical results, including the case of 
$z=2$ quantum critical point, are presented in the supplementary material\cite{SM}.

For the question of cooling from a high temperature to a lower finite temperature, we can apply a similar treatment. However, in this case, $T_f-t^\prime/(\tau k^{zs})\approx T_f$ is not a good approximation, we need to make a higher-order approximation. The specific processing steps are provided in the supplementary material\cite{SM}, and the final result is
\begin{eqnarray}
			\mathcal{D}(t_f)-\mathcal{D}^{\rm th}(T_f)\ \sim
			\begin{cases}  
			\tau^{-\frac{z+1}{zs}}, & z(s-1)\ge1,\\  
			\tau^{-1}, &  z(s-1)<1.
		\end{cases}  \label{fermikzm2}
\end{eqnarray}
Specially, we get a scaling of $\tau^{-1}$ for $s=1$ and $\tau^{-2/3}$ for $s=3$, as shown in Fig. \ref{kzmfermi}.  More numerical results, including the case of 
$z=2$ quantum critical point, are presented in the supplementary material\cite{SM}.

For the bosonic bath, the only difference is $\gamma_k=\gamma_{k+}+\gamma_{k-}$=$\gamma_0\varepsilon_k^s/\tanh[\varepsilon_k/(2T)]$, 
thus for finite $T$, $\gamma_k\sim k^{z(s-1)}$ as $k\rightarrow 0$; therefore,   for a linear heating from a quantum critical point, the scaling is 
\begin{eqnarray}
&&	\mathcal{D}(t_f)-\mathcal{D}^{\rm th}(T_f) \sim 
	\begin{cases}  
		\tau^{-\frac{1}{z(s-1)}}, & z(s-1)\ge1,\\  
		\tau^{-1}, &  z(s-1)<1; 
	\end{cases}  \label{bosekzm1}
\end{eqnarray}
for a linear cooling to a finite temperature, the scaling is 
\begin{eqnarray}
&&		\mathcal{D}(t_f)-\mathcal{D}^{\rm th}(T_f) \ \sim
\begin{cases}  
	\tau^{-\frac{z+1}{z(s-1)}}, & z(s-2)\ge 1,\\  
	\tau^{-1}, &  z(s-2)< 1.
\end{cases}  \label{bosekzm2}
\end{eqnarray}

{\it Free relaxation following a sudden quench.}---The rate  equation (\ref{dynEQ}) is also very suitable for studying the free relaxation behavior after a temperature quench, because the quantum parameters of the system remain unchanged during the relaxation process.
Here, we first consider the simple case of sudden quenches, from which the conclusions derived can be easily adapted to the relaxation processes after linear quenches,  with minor modifications.
 For the case of a sudden quench, i.e., the parameters ($T_i, \mu_i$) is suddenly changed to ($T_f, \mu_f$),  Eq. (\ref{dynEQ}) can be exactly solved, which is written as 
\begin{eqnarray}
	\mathcal{P}_k(t)-\mathcal{P}^{\rm th}_k(\varepsilon_k/T_f)=e^{-2\gamma_kt}\Big[\mathcal{P}_k(0)-\frac{1}{e^{\varepsilon_k/T_f}+1}\Big],    
\end{eqnarray}
where $\varepsilon_k=\varepsilon_k(\mu_f)$, $\gamma_k=\gamma_{k+}+\gamma_{k-}$, and $\mathcal{P}_k(0)$ is the initial state.
 If the quantum parameter $\mu$ remains unchanged during the quench, 
i.e., $\mu_f=\mu_i$, the initial state should be the thermal equilibrium state, 
that is,  $\mathcal{P}_k(0)=1/(e^{\varepsilon_k(\mu_i)/T_i}+1)$, with $T_i$ the initial temperature. However, if the quantum parameter is changed during the quench, additional processing is required for $\mathcal{P}_k(0)$, please refer to the supplementary material for details\cite{SM}.
Integrating over $k$, we obtain the expression of the excess excitation density
\begin{eqnarray}
	\small
	\mathcal{D}(t)-\mathcal{D}^{\rm th}_{T_f}=\frac{1}{2\pi}\int_{-\pi}^{\pi}dke^{-2\gamma_kt}\Big[\mathcal{P}_k(0)-\frac{1}{e^{\varepsilon_k(\mu_f)/T_f}+1}\Big],    \label{mainEQ}
\end{eqnarray}
where $\mathcal{D}^{\rm th}_{T_f}$ is the excitation density of the thermal state at temperature $T_f$.

\begin{figure}[thpb]
	\centering
	\includegraphics[width=1\columnwidth]{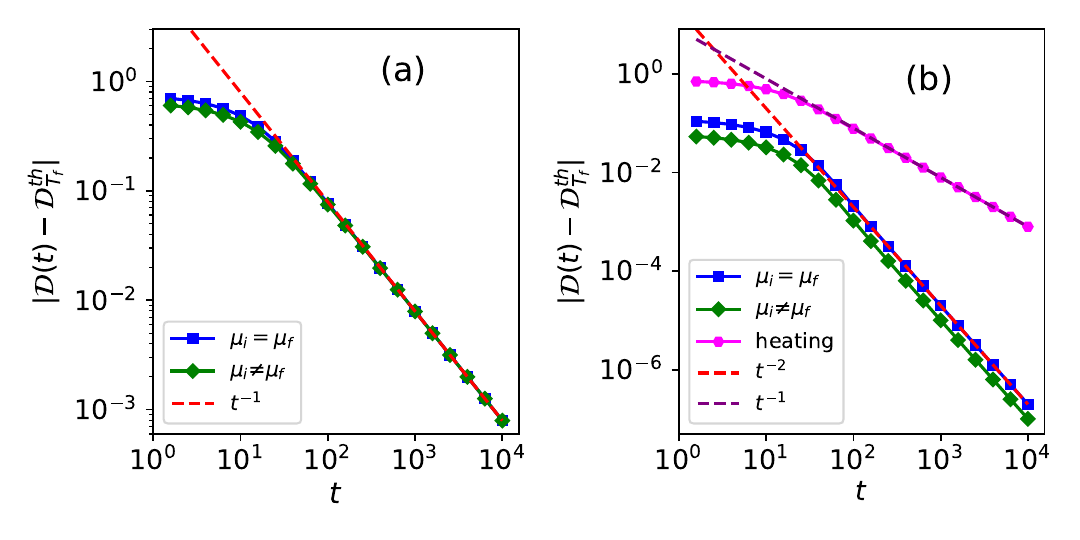}
	\caption{(Color online) 
Free relaxation behaviors of the Kiteav chain (\ref{shortfermion}) following a sudden quench in the fermionic bath with Ohmic spectral density $\mathcal{S}(\varepsilon_k)=\gamma_0\varepsilon_k$:
(a) Relaxation following a sudden quench to the quantum critical point:  ``$\mu_i=\mu_f$'' is the quench from $(T_i, \mu_i)=(5, 1)$ to $(T_f, \mu_f)=(0, 1)$,
``$\mu_i\ne \mu_f$'' is  the quench from $(T_i, \mu_i)=(5, 2)$ to $(T_f, \mu_f)=(0, 1)$.
(b) Relaxation following a sudden quench a finite temperature: 
``$\mu_i=\mu_f$'' is the quench from $(T_i, \mu_i)=(10, 1)$ to $(T_f, \mu_f)=(5, 1)$,
``$\mu_i\ne \mu_f$'' is  the quench from $(T_i, \mu_i)=(10, 2)$ to $(T_f, \mu_f)=(5, 1)$, ``heating'' is the quench from the quantum critical point to (5,1).
The other parameters are set as  $\chi=1$, $\gamma_0=0.01$,  and $L=10^4$.
 }
	\label{scaling}
\end{figure} 
Equation (\ref{mainEQ}) serves as an important starting point for our study of free relaxation behavior. With it, we can readily derive the scaling form of the relaxation behavior after a quench to the quantum critical point.
 We take the  case of fermionic thermal bath with Ohmic spectral density $\mathcal{S}(\varepsilon_k)=\gamma_0\varepsilon_k$ as a typical example.
We consider a simple scenario where $\mu=\mu_c$ is kept  and  the temperature is suddenly cooled to $T_f=0$ (A to C in Fig. \ref{pd}).
In this case, the energy spectrum $\varepsilon_k\rightarrow 0$ in the thermodynamic limit as $k\rightarrow 0$, 
and the main contribution to the integral in Eq. (\ref{mainEQ}) for sufficiently large $t$ comes from the vicinity of the gap-closing point $k=0$;  
in Eq. (\ref{mainEQ}), the term $1/(e^{\varepsilon_k/T_f}+1)=0$ and the energy spectrum can be replaced by the asymptotic form $\varepsilon_k\sim k$, 
then the integration is proportional to $\int e^{\gamma_0 k t} dk$,  thus the scaling form of the excitation density is
$\mathcal{D}(t)\sim t^{-1}$.
This result is demonstrated in Fig. \ref{scaling}(a).  
Furthermore,  for a sudden quench with $\mu_i\ne\mu_f$  to the  quantum critical point, 
the relaxation behavior also fits $t^{-1}$, which is also demonstrated in Fig. \ref{scaling}(a).

More generally, for a thermal bath whose spectral density takes the form $\mathcal{S}(\varepsilon_k)=\gamma_0\varepsilon_k^s$, a sudden cooling to a
quantum critical point with dynamical exponent $z$ (i.e., $\varepsilon_k\propto k^z$) leads to the following  the scaling form of the relaxation behavior  
\begin{eqnarray}
	\mathcal{D}(t)\sim t^{-\frac{1}{zs}}, ~s\ge1. \label{coolfermi}
\end{eqnarray}
The detailed results regarding this question are shown in the supplementary material\cite{SM},
which include the  cases with baths of different spectral densities and also the quenches to the multicritical point ($\mu,\chi$)=(1,0) that has a dynamic exponent of $z$=2.
Furthermore, Eq. (\ref{coolfermi}) is also valid for the quench in a bosonic thermal bath.

We then consider the case of quenching to a finite temperature.
 For a fermionic thermal bath with Ohmic spectral density $\mathcal{S}(\varepsilon_k)=\gamma_0\varepsilon_k$, 
  we consider a quench from $(T_i, \mu_i)$=(10,1) to $(T_f, \mu_f)$=(5,1).
  In this case the  initial state is  $\mathcal{P}_k(0)=1/(e^{\varepsilon_k(\mu_i)/T_i}+1)$,   this term and the term  $1/(e^{\varepsilon_k/T_f}+1)$ 
  can be approximately written as  $\frac{1}{2}(1-k/T_i)$ and  $\frac{1}{2}(1-k/T_f)$, respectively, 
  where we have  replaced  the energy spectrum $\varepsilon_k$ by  the asymptotic form $\varepsilon_k \sim k$.
  Then we find that the integral is proportional to $\int e^{\gamma_0 kt} kdk$, 
  thus  the excitation density $\mathcal{D}(t)$ follows the  asymptotic form $\mathcal{D}(t)-\mathcal{D}_{T_f}^{\rm th}\sim t^{-2}$
when $t$ is large enough.  This result is demonstrated in Fig. \ref{scaling}(b). Furthermore, this result is also valid for describing the relaxation behaviors following a sudden quench with $\mu_i \ne \mu_f$, which is also demonstrated in Fig. \ref{scaling}(b).
However, in Fig. \ref{scaling}(b), we can also see that a sudden quench from the quantum critical point $(T_i,\mu)$=(0,1)  to (5,1) leads to a different relaxation behavior, 
which scales as  $t^{-1}$.  In this case $\mathcal{P}_k(0)=0$, then in Eq. (\ref{mainEQ}), the integral is approximately proportional to  $\int e^{\gamma_0 kt} dk$, 
thus the  scaling form of the long-time limit is $t^{-1}$; 
this result is consistent with Eq. (\ref{coolfermi}), i.e., the case of cooling to a quantum critical point. 
Such situation is special because, in general, the relaxation behavior only depends on the quenched Hamiltonian and has nothing to do with the initial state. 
Here we show that if the initial state is a quantum critical state, then it is an exception. Similar situations appear in the relaxation behavior of isolated quantum systems\cite{I-2023,I-2024,C-2024}.

More generally, for a fermionic thermal  bath whose spectral density takes the form $\mathcal{S}(\varepsilon_k)=\gamma_0\varepsilon_k^s$, 
a sudden heating from a quantum critical point  to a finite temperature  leads to a universal relaxation behavior as
\begin{eqnarray}
\mathcal{D}(t)-\mathcal{D}_{T_f}^{\rm th}\sim t^{-\frac{1}{zs}}, ~ s\ge 1 \label{fermiheating}
\end{eqnarray}
in the long-time limit.  
In contrast, a sudden cooling from a high temperature to a lower finite temperature leads to the following relaxation behavior
\begin{eqnarray}
	\mathcal{D}(t)-\mathcal{D}_{T_f}^{\rm th}\sim t^{-\frac{z+1}{zs}},  ~ s\ge 1. \label{fermiFT}
\end{eqnarray}
We need to pay attention to two special cases.
Firstly, we need to be cautious of a particular scenario: a sudden quench from a quantum critical point $(T=0, \mu_i, \chi_i)$ to $(T>0, \mu_f, \chi_f)$, where $(T=0,\mu_f, \chi_f)$ is also a quantum critical point but with a different gap-closing point compared to  $(T=0, \mu_i, \chi_i)$. In this case, we will obtain the result described by Eq. (\ref{fermiFT}) rather than Eq. (\ref{fermiheating}). The conclusion of Eq. (\ref{fermiheating}) only applies when the gap-closing points are identical before and after the quench, whereas this specific scenario does not meet that criterion.
Secondly, if the post-quench temperature $T_f$ is a finite value but close to zero, the scaling behavior of Eq. (\ref{coolfermi}) will still manifest itself within a certain time interval, which means the relaxation process of the system will experience a crossover from the scaling behavior of Eq. (\ref{coolfermi}) to that of Eq. (\ref{fermiFT})\cite{SM}.

When the question is extended to the bosonic thermal bath, the scaling laws should be modified;
 for the case of heating from the quantum critical point, the scaling law is
 \begin{eqnarray}
 	\mathcal{D}(t)-\mathcal{D}_{T_f}^{\rm th}\sim t^{-\frac{1}{z(s-1)}},  ~ s\ge 2; \label{boseR1}
 \end{eqnarray}
for the case of cooling from a high temperature to a lower finite temperature,  the scaling law is 
\begin{eqnarray}
	\mathcal{D}(t)-\mathcal{D}_{T_f}^{\rm th}\sim t^{-\frac{z+1}{z(s-1)}},   ~ s\ge 2.\label{boseR2}
\end{eqnarray}
The reason is that for a bosonic bath, $\gamma_k=\gamma_{k+}+\gamma_{k-}$=$\gamma_0\varepsilon_k^s/\tanh[\varepsilon_k/(2T)]$, 
thus $\gamma_k\sim k^{z(s-1)}$ as $k\rightarrow 0$.  

\begin{figure}[thpb]
	\centering
	\includegraphics[width=1.0\columnwidth]{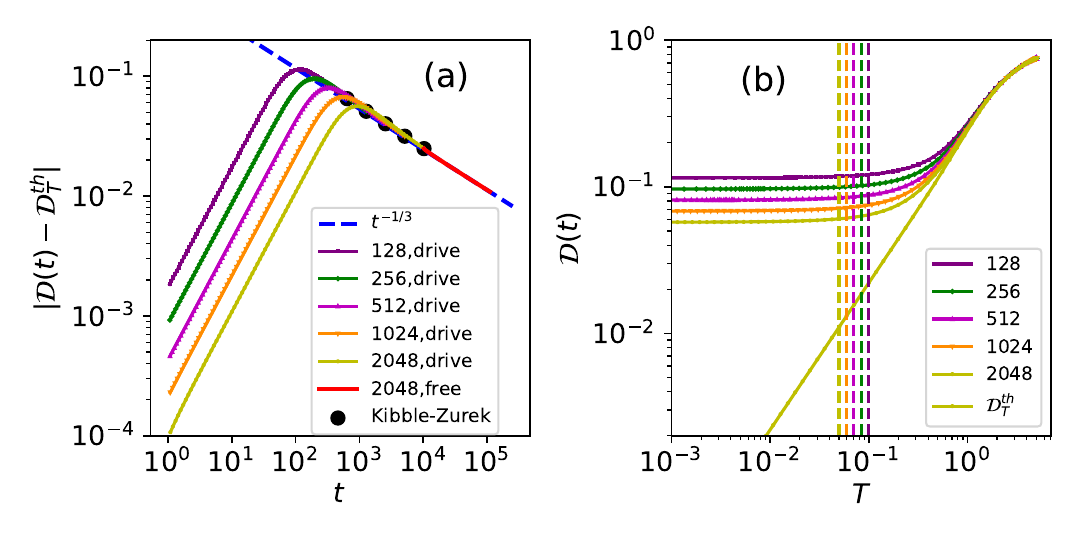}
	\caption{(Color online) 
		Real-time evolution of the  Kiteav chain (\ref{shortfermion}) under linear quench  in a fermionic thermal bath with $s=3$: 
		(a) heating from $T_i=0$ to $T_f=5$, with $T(t)=T_i+t/\tau$; for the case of $\tau=2048$, the free relaxation after the linear driving is also plotted.
		the big black dots indicated the ending points of the linear driving;
		(b) cooling from $T_i=5$ to $T_f=0$, with $T(t)=T_i-t/\tau$; the vertical dashed line indicates the position of  $T(\hat{t}^\prime)$, 
		where $t^\prime=t_f-t$, $t_f=\tau T_i$, and $\hat{t}^\prime\sim \tau^{s/(s+1)}$.
		Other parameters are set as $\mu=\mu_c=1$, $\chi=1$, $\gamma_0=0.01$, and $L=10^4$.}
	\label{lrfermiplot}
\end{figure} 
{\it Real-time dynamics of linear quench.}---The real-time dynamic process of linear quench can be divided into two stages: the driving stage and the free relaxation stage.
A typical example  with $\tau=2048$ is included  in Fig. \ref{lrfermiplot}(a). The scaling behavior in the free relaxation stage is basically consistent with that of a sudden quench, with the only difference being that the time axis needs to be shifted. Taking the case of linear heating starting from the quantum critical point as an example (in fermionic bath), if the ending point of the linear driving is taken as the zero point of time, then the subsequent free relaxation conforms to the following scaling law
\begin{eqnarray}
	\mathcal{D}(t)-\mathcal{D}_{T_f}^{\rm th}\sim (t+t_w)^{-\frac{1}{zs}}, ~ s\ge 1.  \label{fermilinear}
\end{eqnarray}
Here, the critical exponent $1/(zs)$  is exactly the same as that described by Eq. (\ref{fermiheating}) for sudden quench; 
the shifting time $t_w$ is a nonuniversal parameter, which can be obtained through data fitting.
 For Eqs. (\ref{coolfermi}), (\ref{fermiFT}), (\ref{boseR1}) and (\ref{boseR2}), simply replacing $t$ with $t+t_w$ represents the corresponding free relaxation behaviors after linear quench.  For more discussions on this question as well as numerical results, please refer to the supplementary material\cite{SM}.

For the driving stage, we can observe from Fig. \ref{lrfermiplot}(a) that there is a peak for each $\tau$, and the time $\hat{t}$ at which the peak point is located can be obtained by comparing the elapsed driving time with the relaxation time exchanged between the system and the thermal bath 
\begin{eqnarray}
	t \sim \frac{1}{\gamma_0\varepsilon_k^s}\approx \frac{1}{\gamma_0T^s}=\frac{1}{\gamma_0(t/\tau)^s}.\label{R1}
\end{eqnarray}
The reason why the second step holds is due to the fact that the dominant contribution to damping from environment comes from the $\varepsilon_k \sim T$ states\cite{bath5}.
The solution of Eq (\ref{R1}) gives $\hat{t}\sim \tau^{s/(s+1)}$.
Before $\hat{t}$, the exchanging relaxation time is very long, and the response of the system is too slow compared with the driving, causing the difference between 
the transient state of the system and the equilibrium state at the corresponding temperature to widen. This is why we see the difference between  $\mathcal{D}(t)$ and $\mathcal{D}^{\rm th}_{T_f}$ increasing. After $\hat{t}$, the response of the system is relatively fast, and the difference between $\mathcal{D}(t)$ and $\mathcal{D}^{\rm th}_{T_f}$  decreases in a manner close to a power law. Additionally, in current case, for  sufficiently  large $\tau$, the decaying exponent in the driving stage with $t>\hat{t}$  happen to be approximately the same as that in the free relaxation stage. 
This makes the two stages appear to be almost a unified one. However, this is only a  special case. For instance, when cooling down from high temperature to the quantum critical point or a lower finite temperature, the free relaxation process and the driving process are clearly two relatively independent processes\cite{SM}.
It can be observed that $\hat{t}$ resembles the transition point between the frozen and adiabatic stages in the Kibble-Zurek mechanism\cite{KZ3},
in this point, the excess density $\mathcal{D}(\hat{t})$-$\mathcal{D}^{\rm th}_{T_f}\sim \tau^{-1/z(s+1)}$, which is already the same as that of Eq. (\ref{cooluc}).
Equation (\ref{fermikzm1}) is the result of real-time evolution starting from this point and reaching the moment $t=t_f=\tau T_f$.
Therefore, in essence, we do not consider  Eq. (\ref{fermikzm1})) to be the result derived from the standard Kibble-Zurek mechanism.

It is for the above reasons that we call the results of formulas (\ref{fermikzm1}), (\ref{fermikzm2}), (\ref{bosekzm1}) and (\ref{bosekzm2}) Kibble-Zurek-like scaling rather than standard Kibble-Zurek scaling.
In comparing, Eq. (\ref{cooluc}) represents the scenario where the system is linearly cooled from a high temperature to the quantum critical point, which aligns with the standard Kibble-Zurek mechanism\cite{bath5}. In this case,  as shown in Fig. \ref{lrfermiplot}(b),  when the temperature is lowered to  $T(\hat{t}^\prime)\sim \tau^{-1/(s+1)}$ that is near the quantum critical point, the system's response becomes extremely slow, i.e., it is in the freezing stage of  quench,  then the excitation density $\mathcal{D}(t)$ barely changes with the driving, and the final excitation density is determined as $\mathcal{D}(t_f)\approx \mathcal{D}_{T(\hat{t}^\prime)}\sim T(\hat{t}^\prime)^{d/z}=\tau^{-d/z(s+1)}$, i.e., the result of Eq. (\ref{cooluc}). 
Furthermore,  precisely because the system is already in the frozen stage before reaching the end of the driving, the system does not exhibit a power-law decaying stage similar to that seen in the heating process. Consequently, the subsequent free relaxation is  a relatively independent process\cite{SM}.
For the case of cooling from a high temperature to a lower but finite temperature, there is no clear distinction between the frozen and adiabatic stages during the driving stage, specific results can be found in the supplementary material\cite{SM}. 

{\it Conclusions and discussions.}---In summary, taking the quantum Kitaev chain model as an example, we investigated the universal dynamical behaviors induced by quantum criticality under the condition of environmental temperature quench. We found that as long as the quantum parameters are at the critical value, both the excitation density at the ending point of linear quench and the subsequent free relaxation behavior exhibit universal scaling behaviors. The scaling laws manifested by quenching to the quantum critical point at zero temperature and nonzero temperature exhibit different scaling exponents, which are closely related to the dynamical critical exponent of the quantum phase transition.
Generally speaking, these scaling laws are universal and independent of the specific starting point of the quench, with one exception: when the quench starts from the quantum critical point. This is evident in both the Kibble-Zurek-like scaling laws and the scaling laws of free relaxation behavior, as can be seen from the comparison between Eqs. (\ref{fermikzm1}) and (\ref{fermikzm2}), as well as between Eqs. (\ref{fermiheating}) and (\ref{fermiFT}). Similar situations also arise in the problem of free relaxation in isolated quantum systems\cite{I-2023,I-2024,C-2024}.
Quantum criticality plays a crucial role in the universal scaling laws we have discovered.
If the quantum parameter is not at its critical value, the Kibble-Zurek-like scaling behavior in Eq.  (\ref{cooluc}) and all the behaviors of free relaxation will transform into an exponential decaying form\cite{SM}.
However, it is also important to note that power-law dynamical scaling behaviors do not necessarily always originate from quantum criticality; 
 we find that for the quenches  to  finite temperatures, the term $\tau^{-1}$ in Eqs. (\ref{fermikzm1}), (\ref{fermikzm2}),  (\ref{bosekzm1}), and (\ref{bosekzm2}) also exists even when $\mu\ne\mu_c$\cite{SM}, which is due to the last term of Eq. (S5) in the supplementary material \cite{SM}.
 This result is independent of the dynamical critical exponent $z$ and the exponent $s$ of the spectral density of the thermal bath. It is an intrinsic dynamical behavior of  quenching to finite temperatures.

Understanding quantum phase transitions and quantum criticality from the perspective of nonequilibrium and under finite temperature conditions has been a hot topic  in recent years\cite{bath1,bath2,bath3,bath4,bath5,FT1,FT2,FT3,FT4,open-open,nonHermi}. 
Our research offers profound insights into the relationship between quantum criticality and finite-temperature nonequilibrium dynamics from two aspects: Kibble-Zurek-like scaling behavior and free relaxation dynamics. Notably, the Kibble-Zurek-like scaling behavior here is distinct from the standard Kibble-Zurek mechanism. Together with the scaling behavior in the free relaxation process, both aspects pave a new way for us to understand quantum criticality through real-time dynamical behavior, even at finite temperatures.

The research questions explored in this paper are highly extensible, such as exploring quantum quenches by tuning the interaction strength between the system and its environment, and the quench dynamics of physical quantities other than excitation density. Additionally, some quantum systems may exhibit  finite-temperature phase transitions, and the dynamical behaviors resulting from environmental temperature quenches related to these thermal critical points present intriguing questions. Furthermore, although the Kibble-Zurek-like scaling that emerges during linear quench to a finite temperature does not conform to the standard Kibble-Zurek mechanism, we remain hopeful that a modified scaling theory analogous to the Kibble-Zurek mechanism can explain these findings. The research presented here is also worthy of extension to other quantum systems, including open quantum systems with open boundaries\cite{open-open}, non-Hermitian systems\cite{nonHermi}, non-integrable quantum systems\cite{nonInt,nonInt2}, and so forth. Some of these investigations may rely on the latest numerical simulation techniques\cite{FTmethod}.


{\it Acknowledgment.}---C. D. is supported by the National Science Foundation of China (NSFC) under Grant Numbers 11975024 and the Anhui Provincial Supporting Program for Excellent Young Talents in Colleges and Universities under Grant No. gxyqZD2019023.  
L. Z.  is supported by the National Natural Science Foundation of China (No. 12174387), the Chinese Academy of Sciences (Nos. YSBR-057 and JZHKYPT-2021-08), and the Innovative Program for Quantum Science and Technology (No. 2021ZD0302600).

\end{document}


\title{Supplemental Material for\texorpdfstring{\\}{} ``Universal Quench Dynamics of an Open Quantum System''}
	
 \author{Chengxiang Ding}
\email{dingcx@ahut.edu.cn}
\affiliation{School of Microelectronics $\&$ Data Science, Anhui University of Technology, Maanshan 243002, China }
\author{Long Zhang}
\affiliation{Kavli Institute for Theoretical Sciences and CAS Center for Excellence in Topological Quantum Computation, University of Chinese Academy of Sciences, Beijing 100190, China}

	\date{\today}
	
	\maketitle 
	
	\onecolumngrid
	
	\tableofcontents
	
\section{Kibble-Zurek-like scaling, in fermionic bath}
\subsection{Heating from the quantum critical point to finite temperature}
\begin{figure}[thpb]
	\centering
	\includegraphics[width=0.95\columnwidth]{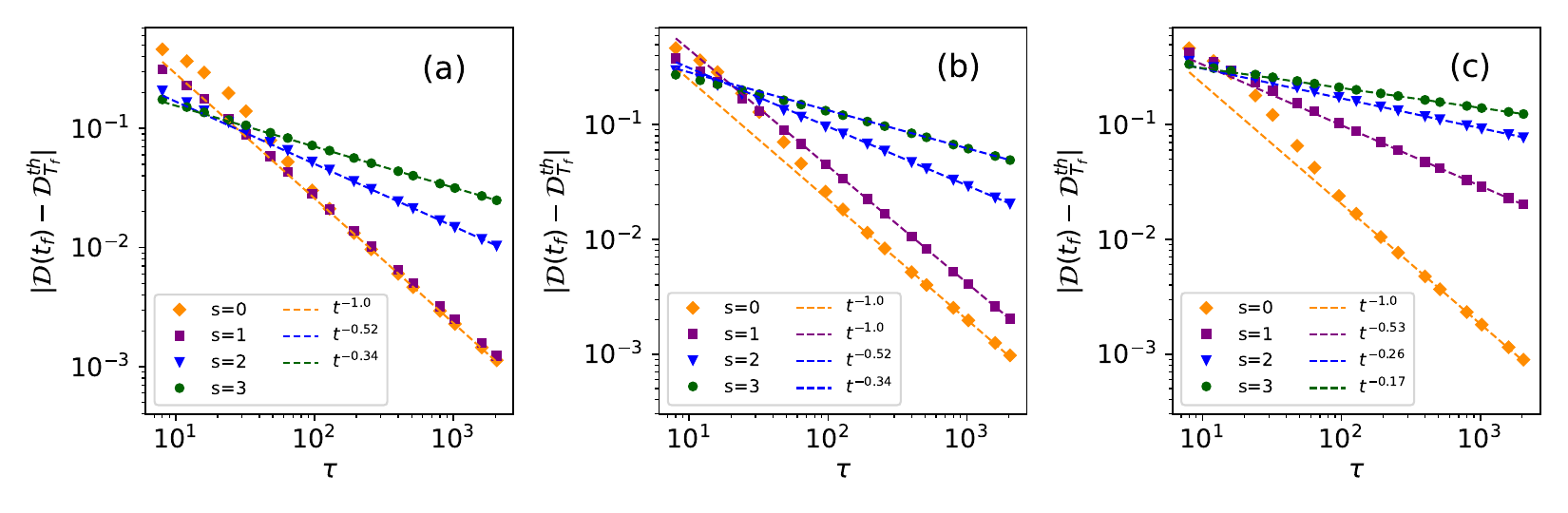}
	\caption{(Color online) Kibble-Zurek-like scaling of the Kitaev chain under a linear quench from the quantum critical point to finite temperature: (a) $(\mu, \chi)=(1,1)$, $z=1$; (b) $(\mu, \chi)=(1,0.5)$, $z=1$; (c) $(\mu, \chi)=(1,0)$, $z=2$.  The  ramping protocol is $T(t)=T_i+t/\tau=t/\tau$,  the ending point of the quench is $T_f=5$, and $t_f=\tau T_f$.
	The system size is $L=10^4$.}
	\label{kzmfermi-heat}
\end{figure}

Here we present more numerical results for the Kibble-Zurek-like scaling behaviors under linear heating from the quantum critical point to finite temperature.
According to the derivation in the main text, in this case the excess excitation density satisfies 
\begin{eqnarray}
	\mathcal{D}(t_f)-\mathcal{D}^{\rm th}(T_f) \sim 
	\begin{cases}  
		\tau^{-\frac{1}{zs}}, & zs\ge 1\\  
		\tau^{-1}, &  zs<1,
	\end{cases} \label{S-fmkz}
\end{eqnarray}
this is exactly Eq. (13) of the main text.
The case of the quantum critical point $(\mu, \chi)=(1,1)$, with dynamical exponent $z=1$,  is shown in Fig. \ref{kzmfermi-heat}(a).
We observe that these numerical results are consistent with the predictions of Eq. (\ref{S-fmkz}). 
We also calculate the case of $(\mu, \chi)=(1,0.5)$ as shown in Fig. \ref{kzmfermi-heat}(b).
In this case, the results are qualitatively  consistent with the case of $(\mu, \chi)=(1,1)$.
This demonstrates that the conclusion of  Eq. (\ref{S-fmkz}) is universal.
The case of the quantum critical point $(\mu, \chi)=(1,0)$, with dynamical exponent $z=2$,  is shown in Fig. \ref{kzmfermi-heat}(c).
In this case, we can also find that the numerical results coincide with the prediction of Eq. (\ref{S-fmkz}).

It should be noted that in numerical calculations, a smaller $\gamma_0$ should be taken for larger $s$ to ensure that the effective interaction $\gamma_0\varepsilon_k^s$ between the system and the environment is sufficiently small, as the Lindblad equation is only applicable to such weakly coupled cases. If the value of $\gamma_0\varepsilon_k^s$ is too large, the results calculated based on the Lindblad equation may be erroneous, especially when $\tau$ is large.

\subsection{Cooling from high temperature to lower finite  temperature}
When the quantum parameter is at the critical value ($\mu=\mu_c$) and the temperature is linearly quenched from a high temperature to a lower finite temperature, 
the excess excitation density of the system at the end of the quenching process also approximately conforms to the Kibble-Zurek-like scaling law. The derivation process is as follows
\begin{eqnarray}
		\mathcal{D}(t_f)=\frac{1}{\pi}\int_{-\pi}^{\pi}\Bigg[ \frac{e^{-2\gamma_kt_f}}{1+e^{\frac{\varepsilon_k}{T_i}}}
	+2\gamma_k\int_{0}^{t_f}\frac{e^{-2\gamma_k(t_f-t)}}{1+e^{\frac{\varepsilon_k}{T_i-t/\tau}}}dt\Bigg]dk\label{D1}
\sim \frac{1}{\pi}\int_{-\pi}^{\pi}\Bigg[\frac{e^{-2\gamma_0k^{zs}t_f}}{1+e^{\frac{k^z}{T_i}}}
	+2\gamma_0k^{zs}\int_{0}^{t_f}\frac{e^{-2\gamma_0k^{zs}(t_f-t)}}{1+e^{\frac{k^z}{T_i-t/\tau}}}dt\Bigg]dk\label{D3}
\end{eqnarray}
where $\gamma_k=\gamma_0\varepsilon_k^s$, and we have taken the approximation $\varepsilon_k\sim k^z$, because the contribution of the integration over $k$ mainly comes from 
the vicinity region of $k\rightarrow0$.  Then by substituting $k^{zs}(t_f-t)=t^\prime$, we get
\begin{eqnarray}
	\mathcal{D}(t_f)\sim\frac{1}{\pi}\int_{-\pi}^{\pi}\Bigg[\frac{e^{-2\gamma_0k^{zs}t_f}}{1+e^{\frac{k^z}{T_i}}}
	+2\gamma_0\int_{0}^{k^{zs}t_f}\frac{e^{-2\gamma_0t^\prime}}{1+e^{\frac{k^z}{T_f+t^\prime/(k^{zs}\tau)}}}dt^\prime\Bigg]dk.
\end{eqnarray}
Taking the following approximation
\begin{eqnarray}
\frac{1}{1+e^{\frac{k^z}{T_f+t^\prime/(k^{zs}\tau)}}}\approx\frac{1}{1+e^{\frac{k^z}{T_f}}}
+\frac{e^{\frac{k^z}{T_f}}}{\big(1+e^{\frac{k^z}{T_f}}\big)^2}\frac{t^\prime}{T_f^2\tau k^{z(s-1)}}
\end{eqnarray}
allows us to first integrate over $t^\prime$, then we get
\begin{eqnarray}
	\mathcal{D}(t_f)\sim &&\frac{1}{\pi}\int_{-\pi}^{\pi}\frac{1}{1+e^{\frac{k^z}{T_f}}}dk
	+\frac{1}{\pi}\int_{-\pi}^{\pi}\Bigg[ \frac{e^{-2\gamma_0k^{zs}t_f}}{1+e^{\frac{k^z}{T_i}}}-\frac{e^{-2\gamma_0k^{zs}t_f}}{1+e^{\frac{k^z}{T_f}}} \Bigg]dk
	-\frac{1}{\pi}\int_{-\pi}^{\pi} \frac{e^{\frac{k^z}{T_f}}}{\big(1+e^{\frac{k^z}{T_f}}\big)^2}\frac{T_i-T_f}{T_f^2}k^ze^{-2\gamma_0k^{zs}\tau T_f}   dk\nonumber\\
	&&-\frac{1}{\pi}\int_{-\pi}^{\pi} \frac{e^{\frac{k^z}{T_f}}}{\big(1+e^{\frac{k^z}{T_f}}\big)^2}\frac{1}{2\gamma_0T_f^2\tau k^{z(s-1)}}e^{-2\gamma_0k^{zs}\tau T_f} dk
	+\frac{1}{\tau}\cdot\frac{1}{\pi}\int_{-\pi}^{\pi} \frac{e^{\frac{k^z}{T_f}}}{\big(1+e^{\frac{k^z}{T_f}}\big)^2}\frac{1}{2\gamma_0T_f^2 k^{z(s-1)}} dk. \label{DD}
\end{eqnarray}
In the right hand of the equation, the first term equals to $\mathcal{D}^{\rm th}(T_f)$,  which is the excitation density of the thermal state at $T_f$.
The other terms describe the asymptotic behavior of $\mathcal{D}(t_f)$ approaching $\mathcal{D}^{\rm th}(T_f)$ as  $\tau$ is large. 
The second, the third,  and the fourth terms  approximately scale as  $\tau^{-\frac{z+1}{zs}}$, for example,  substituting $k\tau^{\frac{1}{zs}} =k^\prime$ in the third term, 
we get
\begin{eqnarray}
\tau^{-\frac{z+1}{zs}}\cdot\frac{1}{\pi}\int \frac{e^{\frac{k^{\prime z}}{T_f \tau^{1/s}}}}{\big(1+e^{\frac{k^{\prime z}}{T_f\tau^{1/s}}}\big)^2}\frac{T_i-T_f}{(z+1)T_f^2}e^{-2\gamma_0k^{\prime zs}T_f}   dk^{\prime z+1}
\approx A(\tau)B\tau^{-\frac{z+1}{zs}} \sim \tau^{-\frac{z+1}{zs}}. \label{Ds3}
\end{eqnarray}
Here $B$ is a constant that does not depend on $\tau$, but 
\begin{eqnarray}
	A(\tau)= \frac{e^{\frac{k^{\prime z}}{T_f \tau^{1/s}}}}{\big(1+e^{\frac{k^{\prime z}}{T_f\tau^{1/s}}}\big)^2}
\end{eqnarray}
is a function of $\tau$, therefore the second and last steps in (\ref{Ds3}) are not exact. However, $A(\tau)$  only weakly 
depends on $\tau$, because $\tau^{1/s}$ is much larger than $k$, therefore we get an approximate scaling as $\tau^{-\frac{z+1}{zs}}$.
Similar question exists in the fourth term of the right hand of Eq. (\ref{DD}). 
For the last term,  it is obvious that it is proportional to $\tau^{-1}$.

From the above analysis, we can conclude that the asymptotic behavior of $\mathcal{D}(t_f)-\mathcal{D}^{\rm th}(T_f)$ for large $\tau$ 
is a mixture of $\tau^{-\frac{z+1}{zs}}$ and  $\tau^{-1}$, i.e., 
\begin{eqnarray}
\mathcal{D}(t_f)-\mathcal{D}^{\rm th}(T_f)\sim C_1(\tau)\tau^{-\frac{z+1}{zs}} + C_2\tau^{-1}, \label{Dres2}
\end{eqnarray}
where $C_1(\tau)$ is a function of $\tau$ but only weakly depend on $\tau$, $C_2$ is constant that does not depend on $\tau$.
When $z(s-1)>1$, the first term decays slower than the second term, thus the asymptotic behavior is dominated by this scaling form.
When $z(s-1)<1$, the second term is the dominating term. In fact even $s=0$, the scaling form is also $\tau^{-1}$; in this case, 
expression of $\mathcal{D}(t_f)$ can be exactly solved\cite{sbath5}.
In summary, we get 
\begin{eqnarray}
	\mathcal{D}(t_f)-\mathcal{D}^{\rm th}(T_f) \ \sim
	\begin{cases}  
		\tau^{-\frac{z+1}{zs}}, & z(s-1)\ge1,\\  
		\tau^{-1}, &  z(s-1)<1.
	\end{cases}  \label{KZFT}
\end{eqnarray}
 
\begin{figure}[thpb]
 	\centering
 	\includegraphics[width=0.75\columnwidth]{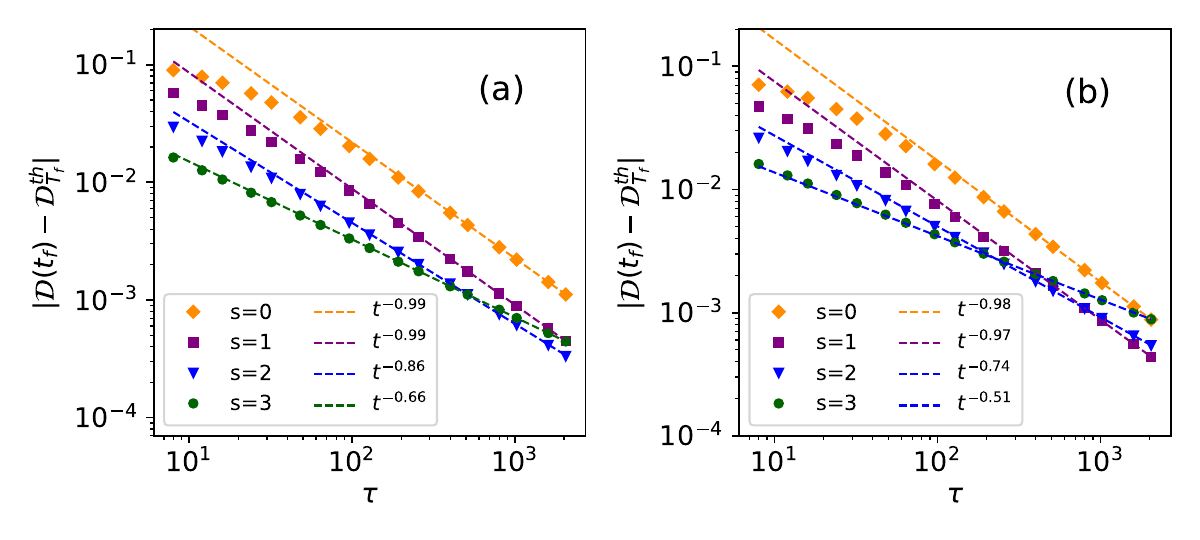}
 	\caption{(Color online) Kibble-Zurek-like scaling of the Kitaev under a linear quench from high temperature to a lower finite temperature: (a) $(\mu, \chi)=(1,1)$, $z=1$; (b) $(\mu, \chi)=(1,0)$, $z=2$.  The  ramping protocol is $T(t)=T_i-t/\tau$,  where $T_i=10$, $T_f=5$,  and $t_f=\tau \cdot(T_i-T_f)$.
 	The system size is $L=10^4$.}
 	\label{kzmfermi-cool}
 \end{figure}

 Figure \ref{kzmfermi-cool}(a) presents some typical examples of this type of quench, from which we can see that the numerical results are in good agreement with the prediction of Eq. (\ref{KZFT}) , except for the case of $s=2$, where there is a slightly larger deviation. In this case, the dependence of $C_1(\tau)$ on $\tau$ in Eq. (\ref{Dres2}) should be relatively stronger, making the crossover process approaching $\tau^{-1}$ significantly longer.  We also studied the case of  $z=2$, as shown in Fig. \ref{kzmfermi-cool}(b).
 In this case, the numerical results are all in good agreement with the predictions of Eq. (\ref{KZFT}).
 
\section{Relaxation following a  quench, in fermionic bath}
So far, all our calculations have been based on the following rate equation\cite{sThirdQ2}
\begin{eqnarray}
	\frac{d\mathcal{P}_k(t)}{dt}=- 2(\gamma_{k+}+\gamma_{k-}) \big[\mathcal{P}_k(t)-\mathcal{P}_k^{\rm th}(\varepsilon_k/T)\big], \label{rate1}
\end{eqnarray}
 which is generally only suitable for cases where the quantum parameters of the system remain unchanged and only the temperature changes. If the quantum parameters also change during the quench, we usually need to use the following dynamical equation\cite{sThirdQ2}
 \begin{eqnarray}
 	&&\frac{d}{dt}\langle a_k^\dagger a_k\rangle = \gamma_{k,1}+\gamma_{k,2}\cos (2\beta_k)-2\gamma_{k,1}\langle a_k^\dagger a_k\rangle 
 	- \varepsilon_k\sin (2\beta_k) \Big(\langle a_k^\dagger a_{-k}^\dagger\rangle^* +\langle a_k^\dagger a_{-k}^\dagger\rangle \Big),\label{Gs1}\\
 	&&\frac{d}{dt}\langle a_k^\dagger a_{-k}^\dagger\rangle=i\gamma_{k,2}\sin (2\beta_k) - 2\gamma_{k,1} \langle a_k^\dagger a_{-k}^\dagger\rangle 
 	+\varepsilon_k \big[ \sin (2\beta_k) (2\langle a_k^\dagger a_k\rangle -1)+2i\cos (2\beta_k) \langle a_k^\dagger a_{-k}^\dagger\rangle \big],\label{Gs2}
 \end{eqnarray}
 which is derived from the method of Third Quantization\cite{sThirdQ2}. 
 Here the Dirac fermions
 \begin{eqnarray}
 	a_k=\frac{1}{\sqrt{L}}\sum_j e^{ijk}c_j,
 \end{eqnarray}
 $\gamma_{k,1}=\gamma_{k+}+\gamma_{k-}$, and $\gamma_{k,2}=\gamma_{k+}-\gamma_{k-}$, where 
 \begin{eqnarray}
 	\gamma_{k+}= \mathcal{S}(\varepsilon_k) f(\varepsilon_k), \quad \gamma_{k-}= \mathcal{S}(\varepsilon_k) [1-\zeta f(\varepsilon_k)],
 \end{eqnarray}
 with $f(\varepsilon_k)=1/(e^{\varepsilon_k/T}+\zeta)$  the Fermi-Dirac distribution (if $\zeta$= +1) or the Bose-Einstein distribution (if $\zeta$= -1) of the bath.
We are interested in the evolution of the excitation number, it can be obtained  by the Bogoliubov transformation
\begin{eqnarray}
	\langle \eta_k^\dagger\eta_k \rangle=\cos (2\beta_k) \Big(\langle a_k^\dagger a_k\rangle-\frac{1}{2}\Big)+\sin (2\beta_k) {\rm Im} [\langle a_k^\dagger a_{-k}^\dagger\rangle]+\frac{1}{2},
\end{eqnarray}
and the Bogoliubov angle $\beta_k$ is defined as
\begin{eqnarray}
	\tan (2\beta_k)=\frac{-\chi sin(k)}{g-\cos(k)}.
\end{eqnarray}
   In the case of a sudden quench, if the quantum parameters do not change, we can use Eq. (\ref{rate1}), and its  solution is 
   \begin{eqnarray}
   	\mathcal{P}_k(t)-\mathcal{P}^{\rm th}_k(\varepsilon_k/T_f)=e^{-2(\gamma_{k+}+\gamma_{k-})t}\Big[\mathcal{P}_k(0)-\frac{1}{e^{\varepsilon_k/T_f}+1}\Big]. \label{rs1}
   \end{eqnarray}
   If the quantum parameters change, we should, in principle, use Eqs. (\ref{Gs1}) and (\ref{Gs2}). 
   However, considering that the quantum parameters only change at the initial moment and remain unchanged afterwards, we can still use the
   rate equation (\ref{rate1}),
   but the step response generated by the sudden change in quantum parameters must be taken into account.
 That is to say, there is a transition in the  initial state of the system from $t=0_-$ to $t=0_+$. For example, when the parameters of the system are suddenly changed from $(T_i, \mu_i, \chi_i)$  to $(T_f, \mu_f, \chi_f)$, the initial thermal state $\mathcal{P}_k(0_-) = 1/(e^{\varepsilon_k/T_i}+1)$ becomes 
 \begin{eqnarray}
\mathcal{P}_k(0_+)=[\cos(2\beta_k^i)\cos (2\beta_k^f)+\sin (2\beta_k^i)\sin (2\beta_k^f) ](\mathcal{P}_k(0_-)-0.5)+0.5,
 \end{eqnarray}
where $\beta_k^i$ and $\beta_k^f$ are the Bogoliubov angles before and after the change of the parameters, respectively.
In this case, the solution of (\ref{rate1}) should be rewritten as 
\begin{eqnarray}
	\mathcal{P}_k(t)-\mathcal{P}^{\rm th}_k(\varepsilon_k/T_f)=e^{-2(\gamma_{k+}+\gamma_{k-})t}\Big[\mathcal{P}_k(0_+)-\frac{1}{e^{\varepsilon_k/T_f}+1}\Big].
\end{eqnarray}
If $\mu_i=\mu_f$ and $\chi_i=\chi_f$, then the question becomes simpler, as $\mathcal{P}_k(0_+)=\mathcal{P}_k(0_-)$.

For the relaxation behavior after linear driving, we can first calculate the driving process according to Eqs. (\ref{Gs1}) and (\ref{Gs2}), and then use the obtained result as the initial state for Eq. (\ref{rate1}). For cases where only the temperature changes, if it is in a fermionic bath, 
the driving process can also be calculated according to Eq. (6) of the main text; 
if it is in a bosonic bath, the integration provided in the supplementary material of Ref. \cite{sbath4} can also be utilized for calculation.

\subsection{Relaxation following a sudden quench}
\begin{figure}[htpb]
	\centering
	\includegraphics[width=1.0\columnwidth]{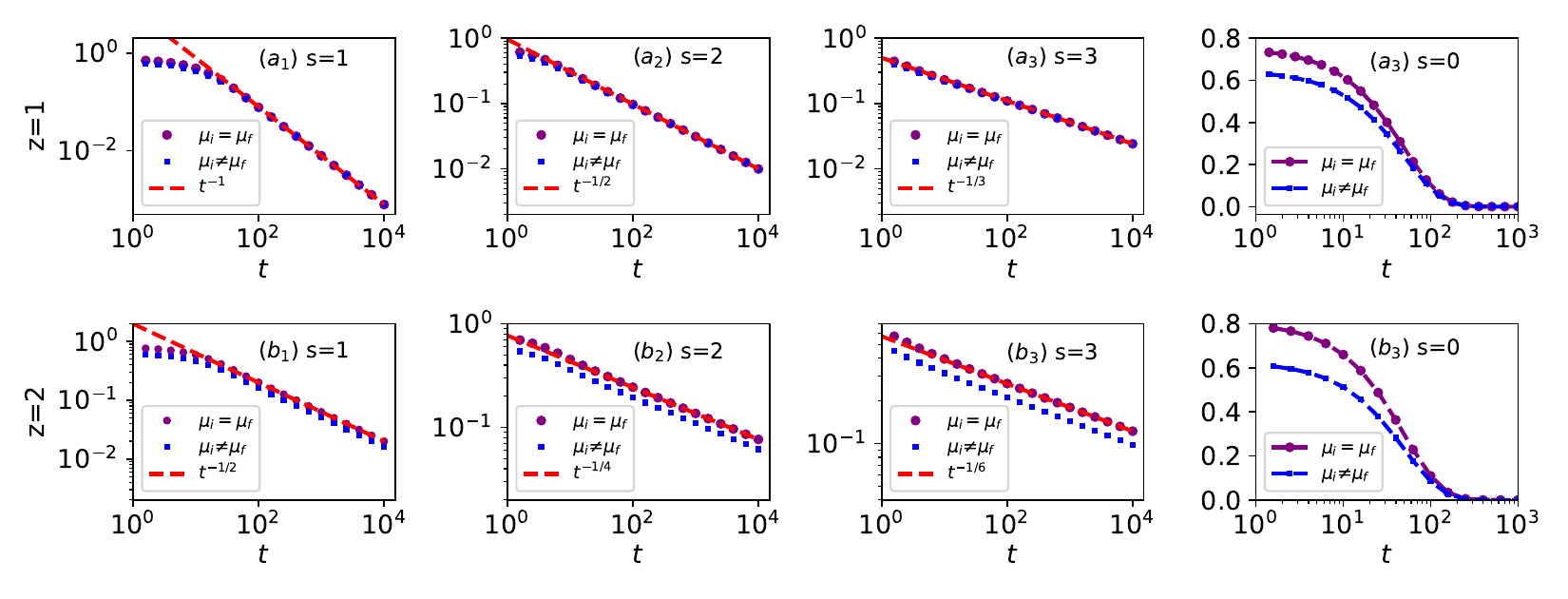}
	\caption{(Color online) Relaxation behaviors of the excess excitation density $|\mathcal{D}(t)-\mathcal{D}^{\rm th}(T_f)|$ of the Kitaev  chain following a sudden quench from high temperature $(T_i=5,\mu_i,\chi)$ to a quantum critical point
		$(T_f=0, \mu_f,\chi)$: the first row shows the results of  the quantum critical point with dynamical exponent $z=1$, i.e., $(\mu_f, \chi)=(1,1)$;
		the second row shows the results of  the quantum critical point with dynamical exponent $z=2$, i.e., $(\mu_f, \chi)=(1,0)$.
		In the case of ``$\mu_i\ne\mu_f$", we set  $\mu_i=2$; the system size is $L=10^4$.}
	\label{relaxcoolQC}
\end{figure} 
Here, we present numerical results of the relaxation behaviors of the quenches from a high-temperature initial  state to the quantum critical point. We consider two representative quantum critical points, namely, the cases of $(\mu, \chi) = (1, 1)$ and $(\mu, \chi) = (1, 0)$, which have dynamical critical exponents $z = 1$ and $z = 2$, respectively. Furthermore, we separately examine scenarios where the quantum parameters change and remain unchanged before and after the quench. All numerical results are shown in Fig. \ref{relaxcoolQC}. We observe that the numerical results are in complete agreement with the predictions of Eq. (19) of the main text. It is noteworthy that the excess excitation density decays exponentially with time in the case of $s = 0$.

\begin{figure}[thpb]
	\centering
	\includegraphics[width=0.75\columnwidth]{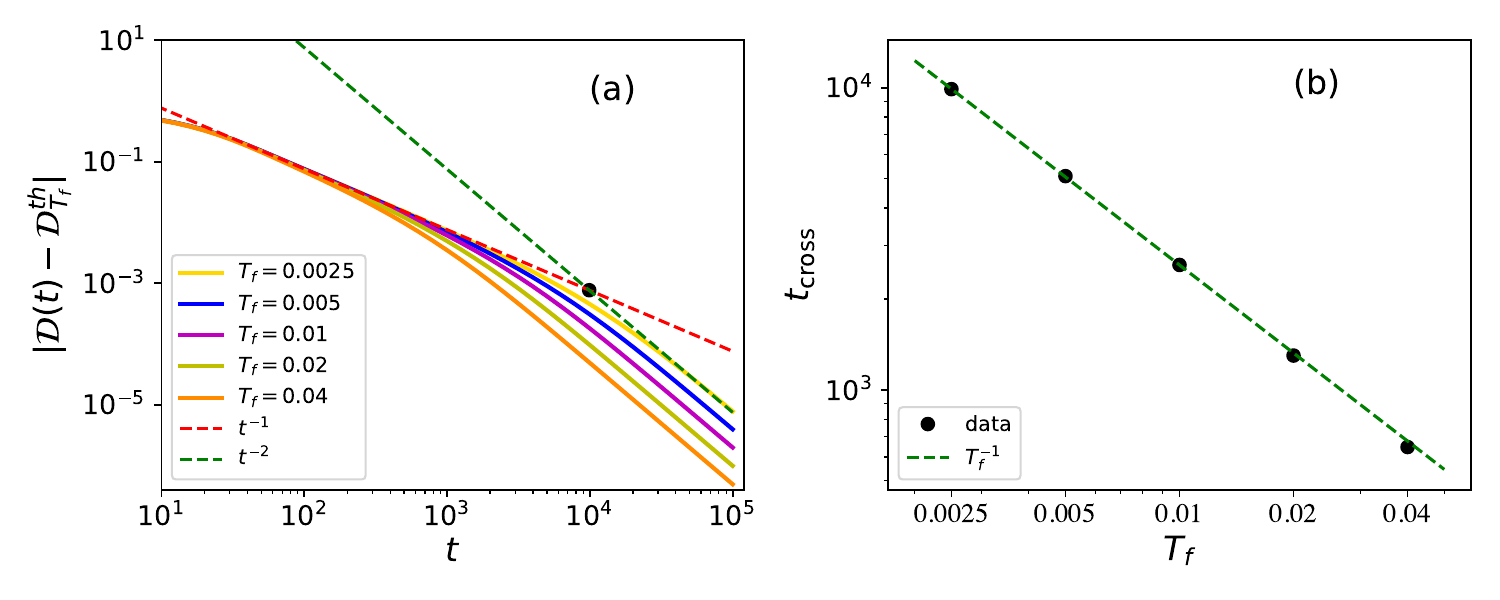}
	\caption{(Color online) Crossover between two different types of scaling in the relaxation behaviors of the excess excitation density $|\mathcal{D}(t)-\mathcal{D}^{\rm th}(T_f)|$ of the Kitaev  chain following a sudden quench from high temperature $(T_i,\mu,\chi)$ 
		to a very low finite temperature $(T_f, \mu,\chi)$, where $(\mu,\chi)=(1,1)$ is a quantum critical point. 
		The other parameters are set as $s=1$, $T_i=5$, and  $L=10^5$.}
	\label{cross}
\end{figure}

For the case of quench to a finite temperature, including the cases of sudden heating from the quantum critical point and cooling from a high-temperature state,
 the numerical results are in line with the predictions of Eqs. (20) and (21) in the main text, respectively. We will not present a plot similar to Fig. \ref{relaxcoolQC} here, instead, we  show the crossover between the  two types of  scaling behaviors when quenching to a temperature that is non-zero but very close to zero.
  A typical result for $s=1$ is shown in Fig. \ref{cross}(a), from which we can see that the scaling is $t^{-1}$ for the time region that is not too long but $t^{-2}$ for the 
  time region that is long enough.   As we demonstrate in the figure for the case of $T_f=0.0025$,
   a crossover time can be defined between the two types of scaling behaviors, 
  as indicated by the big black dot in the figure.  We find that such crossover time scales with $T_f$ as $T_f^{-1}$, as shown in Fig. \ref{cross}(b).
  It is an interesting question to study such crossover scaling in other cases with different $z$ and $s$.

\subsection{Relaxation following a linear quench}
\begin{figure}[thpb]
	\centering
	\includegraphics[width=0.95\columnwidth]{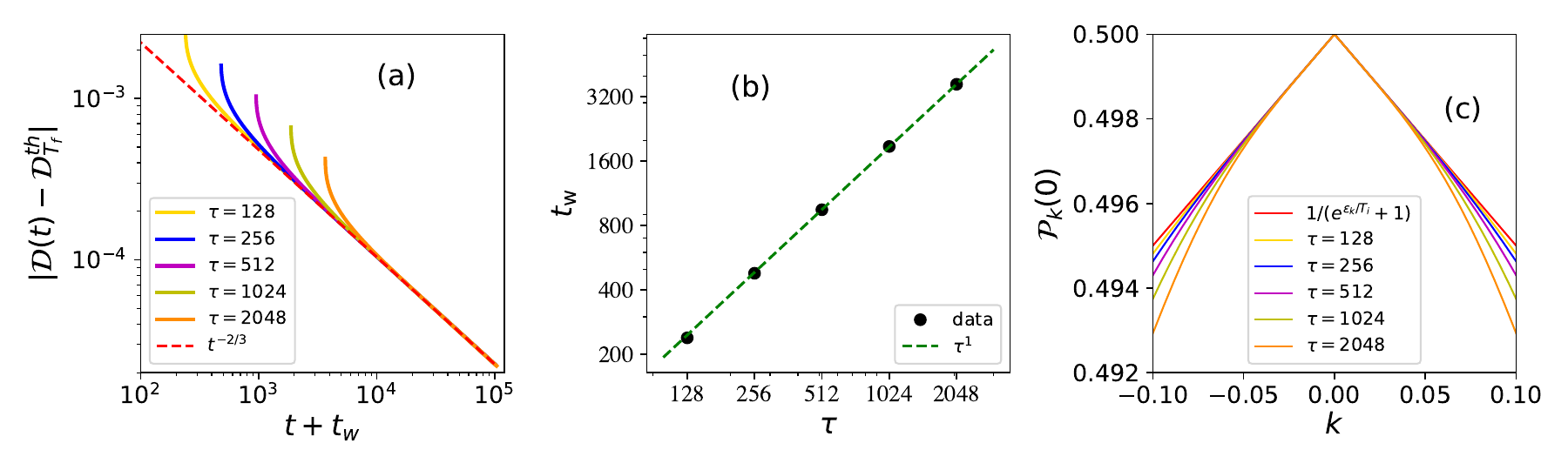}
	\caption{(Color online) (a) Relaxation behaviors of the excess excitation density $|\mathcal{D}(t)-\mathcal{D}^{\rm th}(T_f)|$ of the Kitaev  chain following a linear quench	from $T_i=10$ to $T_f=5$. The ramping protocol is $T(t)=T_i-t/\tau$. However, in the current plots, the zero point of time $t$ is reset to coincide with the ending point of the linear quench. The other parameters are set as $s=3$, $\mu=1$, $\chi=1$, and $L=10^4$. 
		(b) Scaling of the shifting time $t_w$. 
		(c) Asymptotic behaviors of $\mathcal{P}_k(0)$ as $k\rightarrow 0$ for different $\tau$.}
	\label{linearrelax}
\end{figure}

For the relaxation behavior after a linear quench, without loss of generality, we take the case of a cooling down from high temperature to a lower finite temperature as an example. As shown in Fig. \ref{linearrelax}(a), the data can be well fit according to the scaling formula
\begin{eqnarray}
	\mathcal{D}(t)-\mathcal{D}_{T_f}^{\rm th}\sim (t+t_w)^{-\frac{1}{zs}}, {~\rm for ~} s\ge 1.  \label{fermiLFIT}
\end{eqnarray}
In the current example, $z=1$ and $s=3$.  Furthermore, we find that the shifting time $t_w$ scales with $\tau$ as $\tau^{1}$.  
It is an interesting question to study  the scaling of $t_w$ in other cases with different $z$ and $s$.

From Eq. (\ref{rs1}) and the analysis in the main text, we observe that the scaling behavior of the free relaxation process primarily depends on 
the asymptotic behaviors of the initial state $\mathcal{P}_k(0)$ and the thermal state $1/(e^{\varepsilon_k/T_f}+1)$ as $k$ approaches 0. Figure \ref{linearrelax}(c) shows the asymptotic behavior of the state at the endpoint of the linear quench, i.e., $\mathcal{P}_k(0)$, which indicates that it is consistent with the asymptotic behavior of the thermal state $1/(e^{\varepsilon_k/T_i}+1)$ at the corresponding temperature. This is precisely why the relaxation process after linear quench exhibits the same scaling behavior as the corresponding sudden quench. Additionally, we can see that as $\tau$ increases, the range where this asymptotic behavior is consistent with $1/(e^{\varepsilon_k/T_i}+1)$  narrows,  explaining why we need an offset time $t_w$ for larger $\tau$ to obtain better data fitting.

\subsection{Real-time dynamics of the linear driving process}
\begin{figure}[htpb]
	\centering
	\includegraphics[width=0.75\columnwidth]{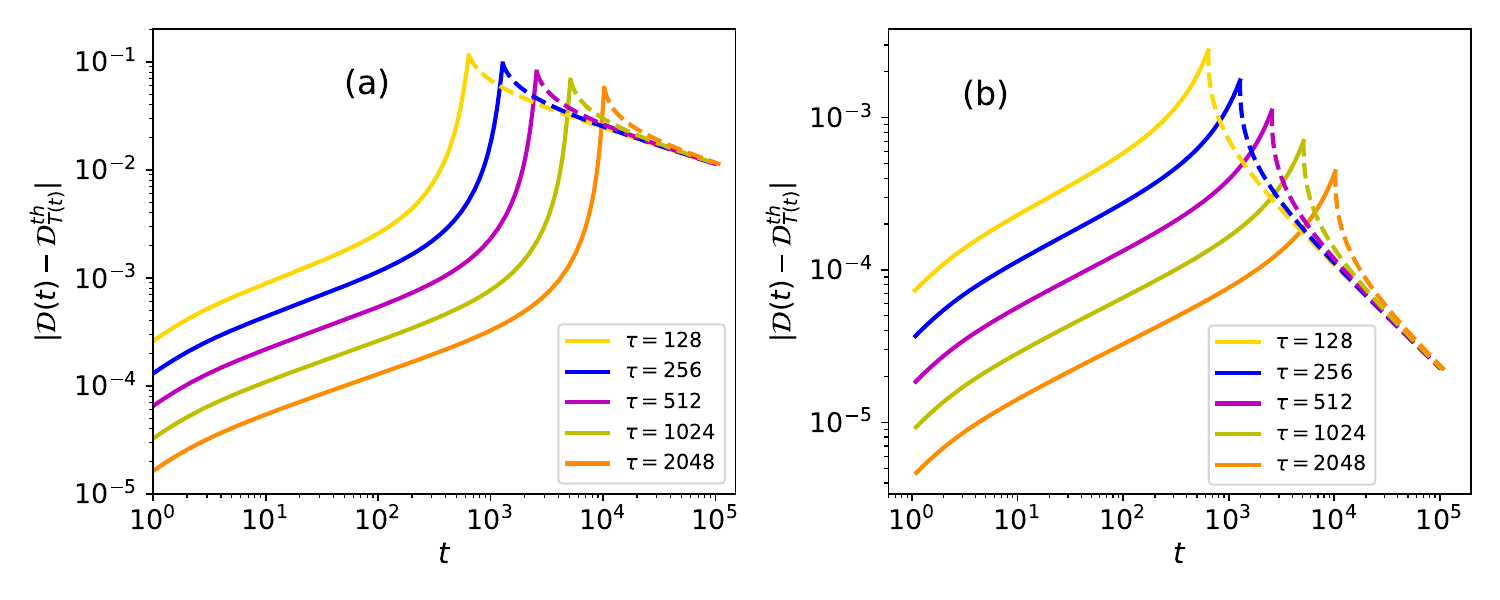}
	\caption{(Color online) (a) Real-time dynamics of  the excess excitation density $|\mathcal{D}(t)-\mathcal{D}^{\rm th}[T(t)]|$ of the Kitaev  chain following a linear quench	from $T_i=5$ to $T_f=0$; the ramping protocol is $T(t)=T_i-t/\tau$.  
		The solid line represents the linear driving stage, while the dashed line represents the free relaxation stage.
		 This is a replot of the Fig. 4(b) of the main text with  the horizontal axis changed as $t$ and 
		the  vertical axis changed as $|\mathcal{D}(t)-\mathcal{D}^{\rm th}[T(t)]|$.
		(b) Real-time dynamics of the  a linear cooling from $T_i=10$ to $T_f=5$.
		The other parameters are set as $s=3$, $\mu=1$, $\chi=1$ and $L=10^4$.}
	\label{evolve}
\end{figure}

In Figure 4(a) of the main text, 
we demonstrate that when linearly quenching from the quantum critical point to a finite temperature, there is a stage during the driving process that approximates a power-law decaying, and the decaying exponent is approximately consistent with that in the subsequent free relaxation process. However, this is a coincidental situation; in general, the linear driving process and the free driving process are two relatively independent processes. For instance, the scenarios shown in Fig. \ref{evolve}, including the linear quench from a finite temperature to the quantum critical point and the linear cooling from a high temperature to a lower finite temperature, illustrate this point. 
We also observe that in the case of quenching from a high temperature to a finite low temperature, the driving process does not exhibit a distinct separation between a frozen stage and an adiabatic stage. This indicates that Eq. (14) in the main text is not a consequence of the standard Kibble-Zurek mechanism.

\section{Quench when quantum parameters do not take critical values, in fermionic bath}
\begin{figure}[thpb]
	\centering
	\includegraphics[width=0.95\columnwidth]{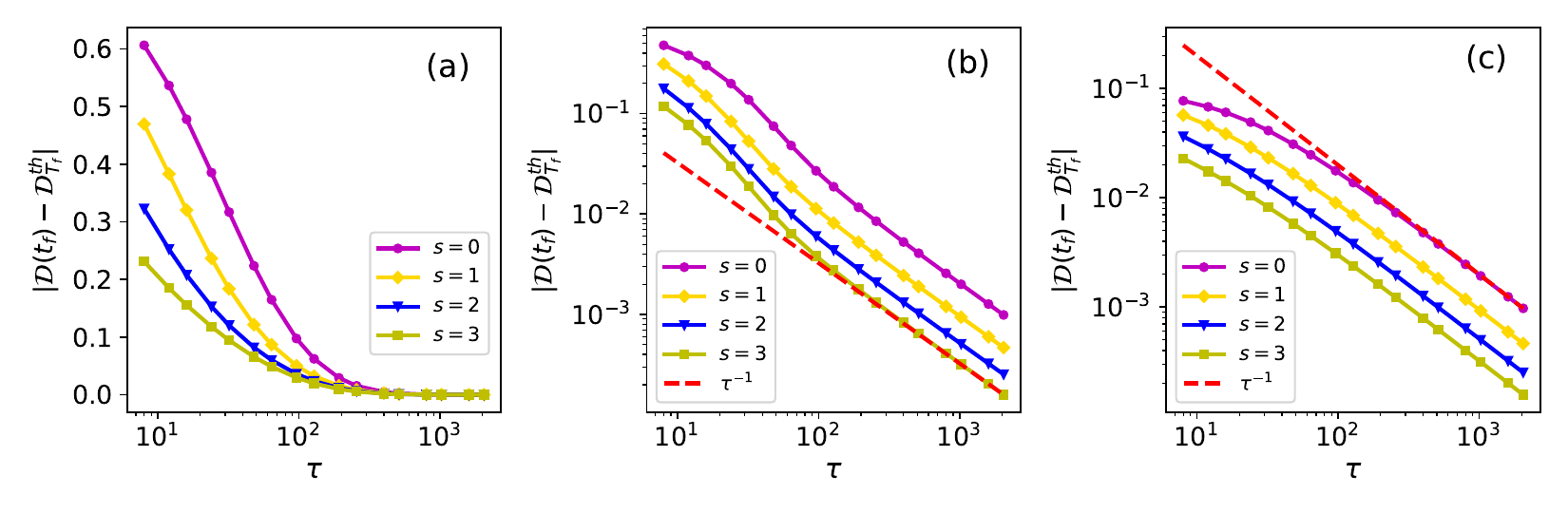}
	\caption{(Color online) Linear quench dynamics of the Kitaev chain when quantum parameters are not at the critical point: 
		(a) quench from $T_i=5$ to $T_f=0$, with ramping protocol $T(t)=T_i-t/\tau$;
		(b) quench from $T_i=0$ to $T_f=5$, with ramping protocol $T(t)=T_i+t/\tau$;
		(c) quench from $T_i=10$ to $T_f=5$, with ramping protocol $T(t)=T_i-t/\tau$.
		Here $t_f=\tau|T_i-T_f|$ is the time of the ending point of the quench.
		The other parameters are set as $\mu=0.5$, $\chi=1$, and $L=10^4$.
	Please note that (a) is a quasi-log plot, while (b) and (c) are log-log plots.}
	\label{nonQCkzm}
\end{figure}

When the quantum parameters are not at their critical values, for the case of linearly cooling from high temperature to the quantum critical point, the scaling behavior described in Eq. (5) of the main text transforms into an exponential decaying behavior, as shown in Fig. \ref{nonQCkzm}. However, quench to a finite temperature point still leads to a Kibble-Zurek-like scaling, but its form is always $\tau^{-1}$. This scaling behavior actually originates from the last term in Eq. (\ref{DD}), which is an intrinsic universal dynamical scaling behavior that does not depend on the specific dynamical critical exponent $z$ or the spectral density exponent $s$ of the bath.

\begin{figure}[thpb]
	\centering
	\includegraphics[width=0.95\columnwidth]{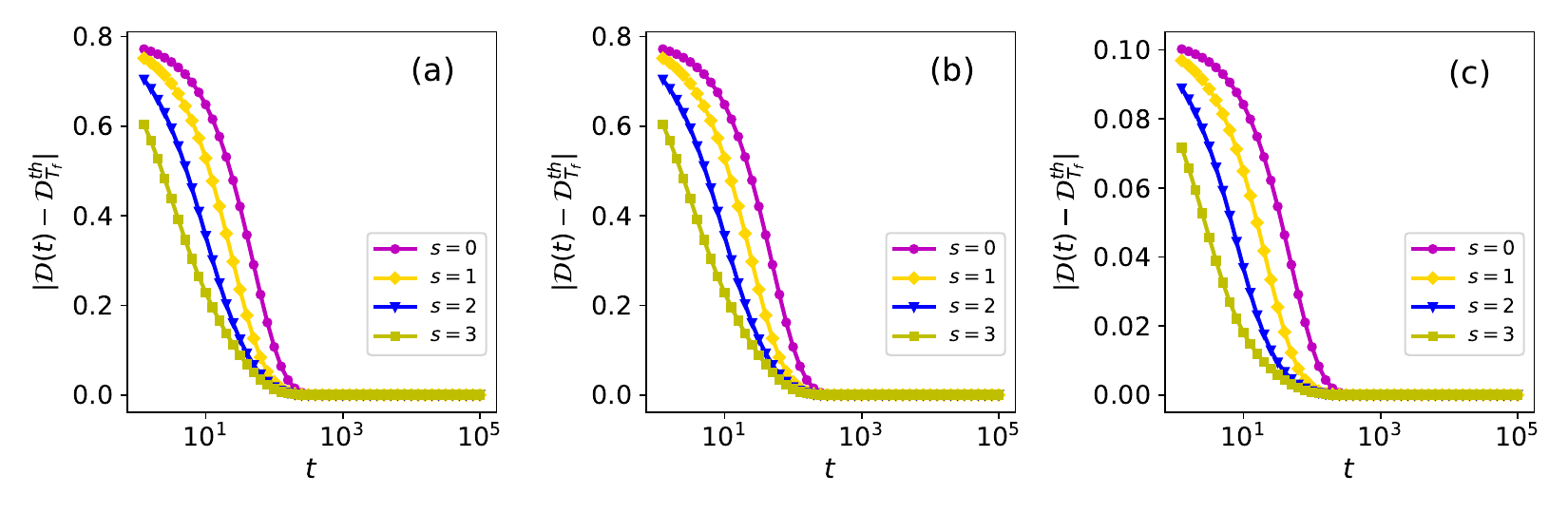}
	\caption{(Color online) Free relaxation of the Kitaev chain when quantum parameters are not at the critical point: 
		(a) sudden quench from $T_i=5$ to $T_f=0$ ;
		(b) sudden quench from $T_i=0$ to $T_f=5$;
		(c) sudden quench from $T_i=10$ to $T_f=5$.
		The other parameters are set as $\mu=0.5$, $\chi=1$, and $L=10^5$.}
	\label{nonQCrelax}
\end{figure}

\begin{figure}[thpb]
	\centering
	\includegraphics[width=0.75\columnwidth]{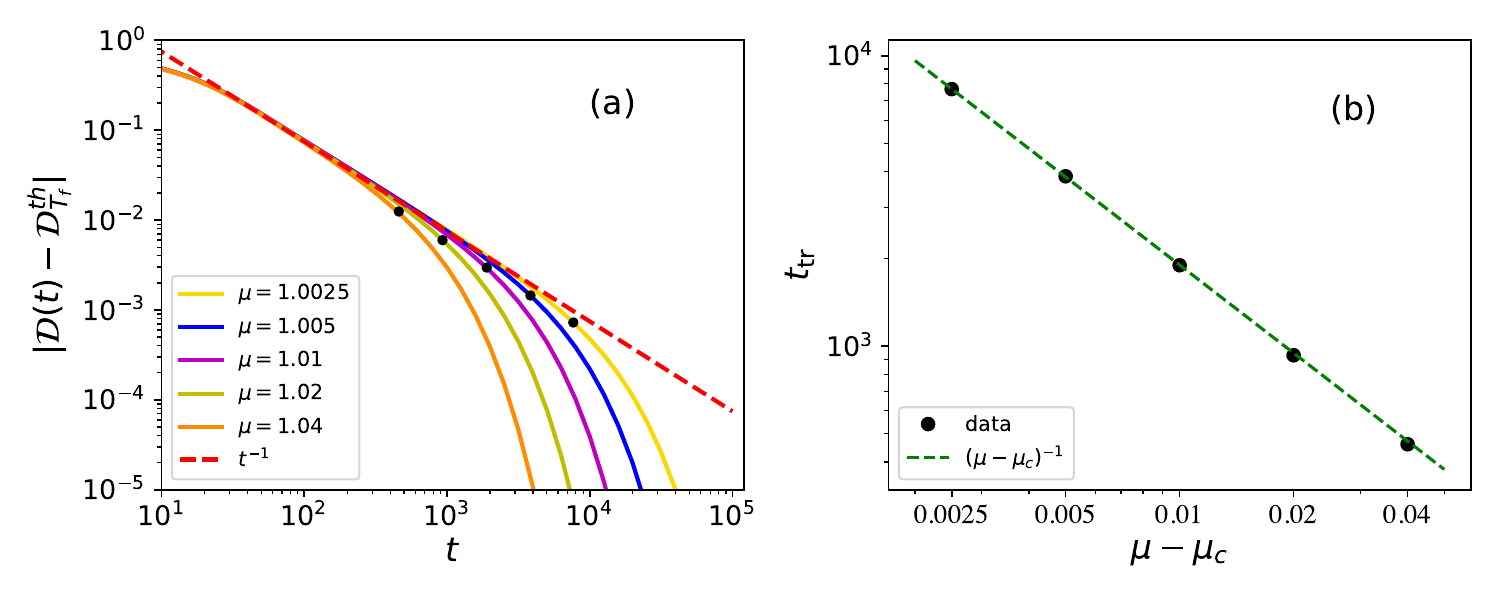}
	\caption{(Color online) (a) Free relaxation of the Kitaev chain under a sudden quench when quantum parameters are not at the critical point, where $s=1$, $\chi=1$, and $L=10^5$. (b) the transition time $t_{\rm tr}$ scales with the $\mu$ as $(\mu-\mu_c)^{-1}$, where $\mu_c=1$.}
	\label{nonQCcross}
\end{figure}

For the relaxation process, as long as the quantum parameters do not take critical values, all the scaling behaviors of power-law decaying we discovered will turn into exponential decaying forms, as shown in Fig. \ref{nonQCrelax}. However, if the quantum parameters take values that are not equal to but very close to the critical point, the excess excitation density can still exhibit a power-law decaying form within a certain interval of time. Figure \ref{nonQCcross}(a) shows a typical example. Nevertheless, when the time is sufficiently long, the real-time dynamical behavior will eventually change to an exponential decaying form; 
 the transition point from power-law decay to exponential decay, indicated by the large black dots in Figure \ref{nonQCcross}(a),  follows a scaling law of the form $t_{tr} \sim (\mu-\mu_c)^{-1}$, as illustrated in Fig. \ref{nonQCcross}(b).
Technically, this transition time can be defined as follows: Set a small quantity $\delta$, compare the value of $|\mathcal{D}(t)-\mathcal{D}^{\rm th}_{T_f}|$ when $\mu>\mu_c$ with that when $\mu=\mu_c$, the corresponding time when the difference between the two equals $\delta$ is defined as the transition time $t_{\rm tr}$.
It is an interesting question to study such crossover scaling in other cases with different $z$ and $s$.
